%% file: tpa_timing.tex
\documentclass[fleqn,usenatbib]{mnras}

\usepackage{newtxtext,newtxmath}
\usepackage[T1]{fontenc}

\DeclareRobustCommand{\VAN}[3]{#2}
\let\VANthebibliography\thebibliography
\def\thebibliography{\DeclareRobustCommand{\VAN}[3]{##3}\VANthebibliography}

\usepackage{graphicx}	% Including figure files
\usepackage{amsmath}	% Advanced maths commands
\title[Timing of 597 pulsars]{The Thousand-Pulsar-Array programme on MeerKAT XIII: Timing, flux density, rotation measure and dispersion measure timeseries of 597 pulsars}

\author[M.~J. Keith et al.]{M.~J. Keith$^{1}$\thanks{E-mail: mkeith@pulsarastronomy.net},
S.~Johnston$^{2}$,
A.~Karastergiou$^{3}$,
P.~Weltevrede$^{1}$,
M.~E. Lower$^{2}$,
A.~ Basu$^{1}$,
B.~Posselt$^{3}$,\newauthor
L.~S.~Oswald$^{3,4}$, 
A.~Parthasarathy$^5$,
A.~D.~Cameron$^{6,7}$,
M.~Serylak$^{8,9}$ and
S.~Buchner$^{10}$
\\
$^1$Jodrell Bank Centre for Astrophysics, Department of Physics and Astronomy, University of Manchester, Manchester M13 9PL, UK\\
$^2$Australia Telescope National Facility, CSIRO Space and Astronomy, PO~Box~76, Epping NSW~1710, Australia\\
$^3$Department of Astrophysics, University of Oxford, Denys Wilkinson Building, Keble Road, Oxford OX1 3RH, UK\\
$^{4}$Magdalen College, University of Oxford, Oxford OX1 4AU, UK\\
$^5$Max-Planck-Institut für Radioastronomie, Auf dem Hügel 69, D-53121 Bonn, Germany\\
$^6$Centre for Astrophysics and Supercomputing, Swinburne University of Technology, PO Box 218, VIC 3122, Australia\\
$^7$ARC Centre of Excellence for Gravitational Wave Discovery (OzGrav), Swinburne University of Technology, PO Box 218, VIC 3122, Australia\\
$^8$SKA Observatory, Jodrell Bank, Lower Withington, Macclesfield, SK11 9FT, UK\\
$^9$Department of Physics and Astronomy, University of the Western Cape, Bellville, Cape Town, 7535, South Africa\\
$^{10}$South African Radio Astronomy Observatory, 2 Fir Street, Black River Park, Observatory 7925, South Africa
}

\date{Accepted XXX. Received YYY; in original form ZZZ}

\pubyear{2023}

\begin{document}
\label{firstpage}
\pagerange{\pageref{firstpage}--\pageref{lastpage}}
\maketitle

\begin{abstract}
We report here on the timing of 597 pulsars over the last four years with the MeerKAT telescope. We provide Times-of-Arrival, pulsar ephemeris files and per-epoch measurements of the flux density, dispersion measure (DM) and rotation measure (RM) for each pulsar. In addition we use a Gaussian process to model the timing residuals to measure the spin frequency derivative at each epoch. We also report the detection of 11 glitches in 9 individual pulsars. We find significant DM and RM variations in 87 and 76 pulsars respectively. We find that the DM variations scale approximately linearly with DM, which is broadly in agreement with models of the ionised interstellar medium. The observed RM variations seem largely independent of DM, which may suggest that the RM variations are dominated by variations in the interstellar magnetic field on the line of sight, rather than varying electron density. We also find that normal pulsars have around 5 times greater amplitude of DM variability compared to millisecond pulsars, and surmise that this is due to the known difference in their velocity distributions.
\end{abstract}


%%%%%%%%%%%%%%%%%%%%%%%%%%%%%%%%%%%%%%%%%%%%%%%%%%

%%%%%%%%%%%%%%%%% BODY OF PAPER %%%%%%%%%%%%%%%%%%

\section{Introduction}
The radio emission from pulsars is thought to originate from close to the surface of the star and locked to its rotation. The radio emission is usually highly linearly polarised and is beamed along the magnetic axis, resulting in pulses of emission being detected by a distant observer. The time-of-arrival of the pulses at an Earth-based observatory yields much information, it tells us the rotation rate of the pulsar, how fast it slows down, how irregular the slow-down is and about transient events such as glitches. Together, these inform us about the evolution of pulsars over time and provides insight into the composition and structure of the neutron star interior. Additionally the traverse of the emission through the interstellar medium (ISM) gives information on the composition, structure and magnetic fields of the ISM via measurements of dispersion measure (DM, e.g. \citealp{ymw17}) and rotation measure (RM, e.g. \citealp{hmvd18}).

The Thousand Pulsar Array (TPA) programme on the MeerKAT telescope \citep{tpa20} forms part of the larger MeerTime key science project \citep{mtime}. The top-level goals of the TPA are two-fold. First to use the full sensitivity of the MeerKAT telescope to provide a census of both the integrated and single-pulse properties for more than 1000 pulsars \citep{pjk+21,rvm23,census23,song23}. 
The second major goal is to investigate the time-varying properties of pulsars through regular, high fidelity \citep{song21} observations of a large ($\sim\!600$) and diverse sample of pulsars. These observations have a median cadence of 27 days, though the exact observing pattern has changed several times over the course of the programme primarily due to scheduling constraints on the telescope. 

This paper describes the data products from the TPA regular monitoring programme that are published alongside this paper. This first data release consists of time of arrivals (ToAs) and pulsar ephemerdies for 597 pulsars, as well as measurements of spin-down rate, flux density, rotation measure and dispersion measure for each pulsar over the period 2019 March to 2023 May.
Pulse profile shapes are also monitored within the TPA, details of which will appear in a subsequent paper, with initial results published in Basu et al. (2023; submitted).

Section~\ref{observations_and_pulsars} briefly outlines the observations and defines the sample of pulsars in the data release. Section~\ref{method} describes the methods used to measure each of the parameters presented in the tables and figures comprising the data release, which are themselves described in Section~\ref{tables_and_figures}.
Some broad highlights of the data are presented in Section~\ref{highlights}, with a particular focus on the measurements of DM and RM variations as an example of what may be done using the data.

\section{The TPA data}
\label{observations_and_pulsars}
\subsection{Observations}
Observations for the TPA programme fall into two categories. For the census projects, observations were carried out with at least 60 antennas of the MeerKAT array. Observations were done with the L-band receiver and, generally, every pulsar was observed once with an integration time determined using the prescription outlined in \citet{song21}. For the monitoring project we aim to observe some 500 pulsars once per month. These observations split the array into two sub-arrays each consisting of $\sim30$ antennas. Here, observation times are very short (typically less than 2~mins in duration) in order to observe as many pulsars as possible in the monthly allocation of 16~hours of telescope time. The data in this data release combine both types of observations, though the vast majority of data values come from the regular observations.
Although we aim for continuous observations with a regular sampling, around half of the pulsars exhibit a gap in observations due to periods where pressure on available telescope time required us to reduce the number of pulsars observed.
Hence, there is a gap of around one year, from 2021 March to 2022 February, for 294 pulsars, and a gap of around 200 days, from 2019 Nov to 2021 May, for 74 pulsars.

A description of the observing system can be found in \citet{mtime} and \citet{tpa20}. In brief, we used the observational band from 896 to 1671~MHz with 928 frequency channels. The data are folded into sub-integrations each of length $8\,\mathrm{s}$ for the duration of the observation and there are 1024 phase bins per pulse period. Polarization calibration follows the method outlined in \citet{sjk+21} and flux calibration is achieved as described in \citet{census23}. Data outputs are in {\sc psrfits} format \citep{hvm04}. In total, more than 24,000 observations have been made for the TPA programme over the 4-yr timespan.

\subsection{The pulsar sample}

\begin{figure}
\begin{center}
\includegraphics[width=8cm,angle=0]{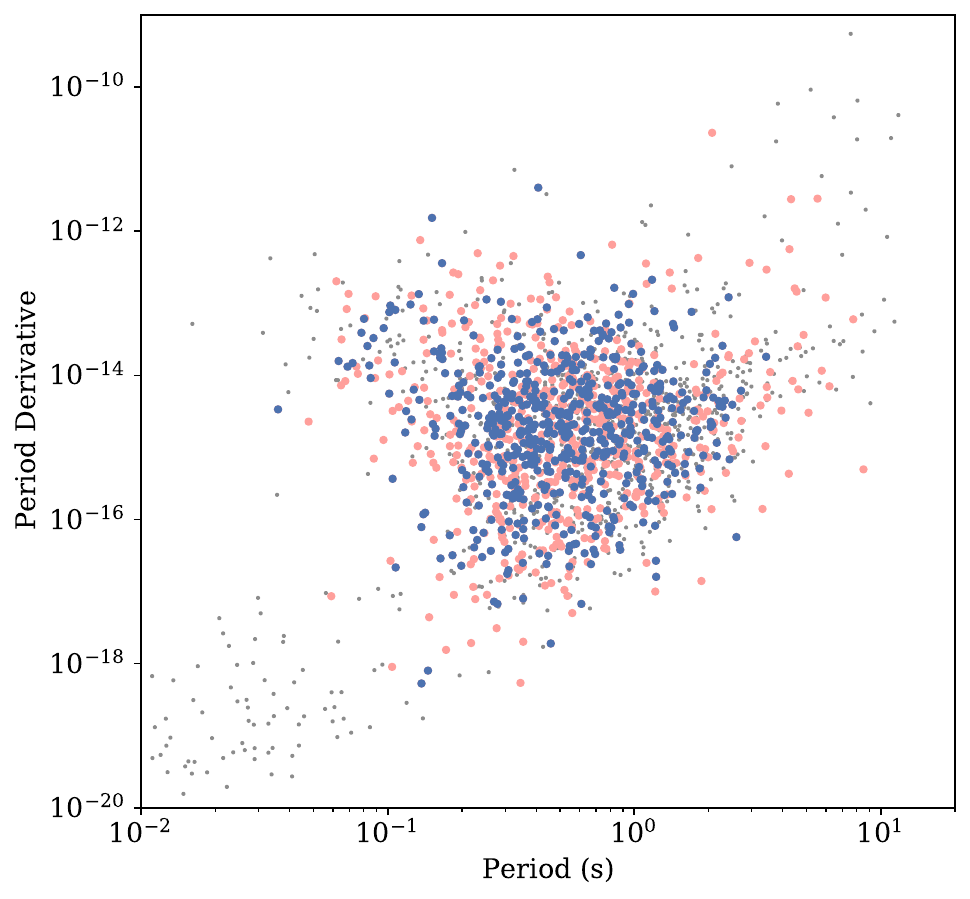} \\
\end{center}
\caption{Period vs period derivative ($P$-$\dot{P}$ diagram) for the 597 pulsars in this data release (blue), the full TPA census (pink) and all pulsars in the ATNF pulsar catalogue (\textsc{psrcat}, version 1.70, \url{https://www.atnf.csiro.au/research/pulsar/psrcat}). Note that not all pulsars in $\textsc{psrcat}$ are shown due to the plot axis limits.}
\label{fig:ppdot}
\end{figure}

\begin{figure*}
\begin{center}
\begin{tabular}{c}
\includegraphics[width=16cm,angle=0]{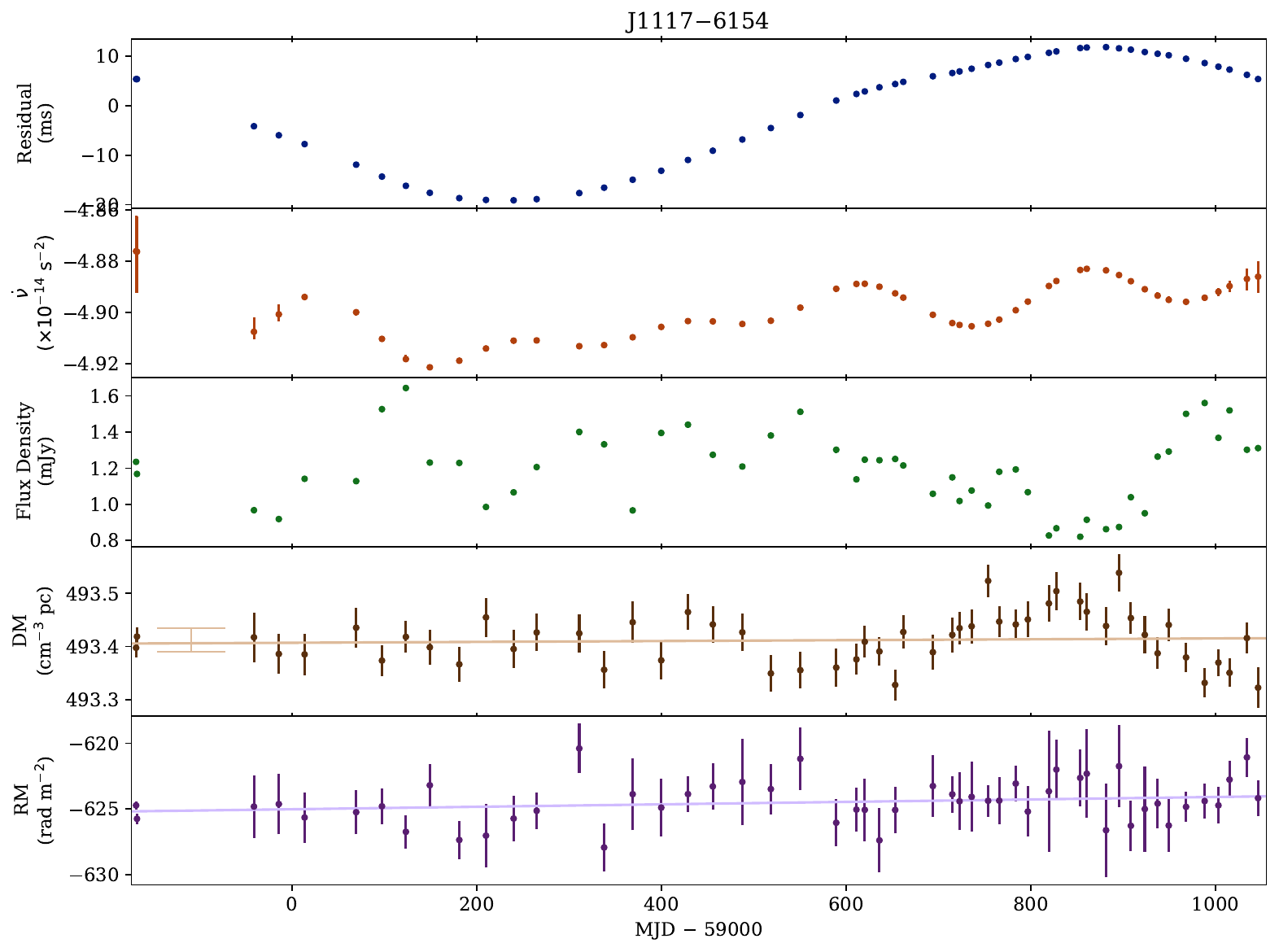} \\
\end{tabular}
\end{center}
\caption{Top panel shows the timing residuals as a function of epoch for PSR PSR~J1117$-$6154, followed by the values of $\dot{\nu}$, the flux density, the DM, and ionosphere corrected RM. The solid lines show the best fitting slopes to the DM and RM versus time. 1-$\sigma$ errors are shown on all panels, but too small to be visible on residual and flux density. The large error bar on the DM fit represents an estimate of the systematic error in the mean DM due to frequency evolution of the pulse profile (see Section~\ref{SecDMvsTime} for a description of how this is estimated).}
\label{fig:bigplot}
\end{figure*}

There are in excess of 1250 pulsars observed as part of the TPA \citep{census23,song23}.
Many of these only have a one or two census-style observations, but there are 614 pulsars that had at least 10 observations in the period from 2019 March to 2023 May.
Pulsars were selected for the regular monitoring campaign based on the quality of data that could be obtained in the very short observation time, and the list was refined several times with the priority on ensuring data of good quality whilst also trying to keep a good spread of pulsars over the $P$-$\dot{P}$ diagram (see Figure \ref{fig:ppdot})
The latter criteria was included to reduce selection bias when the data are used for studies of properties of the pulsar population, though we note that the TPA does not observe millisecond pulsars (MSPs) and the MeerTime programme to observe MSPs is described elsewhere \citep{mtimemsp}. 
From these 614 we exclude the 7 pulsars which have binary companions (PSRs~J0045$-$7319, J0823+0159, J1141$-$6545, J1302$-$6350, J1740$-$3052, J1906+0746 and J1930$-$1852) as these require more careful analysis and will be described elsewhere. We further exclude three pulsars (PSRs~J1717$-$4054, J1906+0414 and J1929+1357) which have a high nulling fraction and hence very few detections, and one pulsar (PSR~J1644$-$4559) which was used mainly for system tests. Finally, there are six pulsars (PSRs~J0835$-$4510, J1301$-$6305, J1420$-$6048, J1614$-$5048, J1718$-$3825 and J1907+0631) which we exclude as we were unable to obtain a coherent timing solution because of the presence of large glitches.
There are therefore a total of 597 pulsars presented in this paper.

\section{Measurement methodology}
\label{method}
\subsection{Timing}
\label{timing}
\subsubsection{Templates and ToA measurement}
\label{SectTemplatesAndToAs}
ToA measurements are one of the data products from the the MeerPipe pipeline which runs automatically on all TPA observations.
The ToAs are computed on the 8 sub-band, fully time-averaged, data using \textsc{psrchive} to cross correlate the observed pulse profile with a noise-free template, using the Fourier-domain markov-chain (FDM) method in \textsc{pat} \citep{hvm04}.
A single template is used across all 8 sub-bands, generated from the `best' TPA observation, which is typically the observation from the TPA census, which uses the full MeerKAT array and much longer intergration time than the typical TPA timing observations, or from averaging several observations where signal-to-noise ratio is low.
For most pulsars the noise-free template is generated by means of a Gaussian process (GP) modelling of the observed average pulse profile \citep{jk19}.
To improve the performance of the template, the profile is tapered off to zero outside the on-pulse window computed in \citet{pjk+21}, which ensures that no information from the off pulse is used in the template matching.
Each automated template was inspected and for 58 pulsars, notably those with scattering tails, interpulses or very complex shapes, the automated template generation was deemed unsuitable. For these pulsars, templates were computed by fitting von Mises functions\footnote{Functions with the form of the unnormalied von Mises distribution, $f(\phi)~=~\exp(-\cos(\phi))$.} `by hand' using \textsc{paas} from \textsc{psrchive}, or in the case of two highly scattered pulsars (PSRs J1701$-$4533 and J1828$-$1101), using von Mises functions convolved with a single exponential with a bespoke routine.
Since only a single template was used across the whole band, users should be aware that the resulting ToAs will include effects of profile evolution which must be accounted for in any later timing analysis.

Initial timing was performed using \textsc{tempo2} to compute and fit the timing residuals.
The initial ephemeris was taken from \textsc{psrcat}, and coarsely refined whilst searching for possible phase wraps between observing epochs.
For some pulsars with significant spin noise or glitch activity, the catalogue ephemerides lost phase at a rate of up to several turns per day, so for some cases we relied on ephemerides provided by long-term timing programmes with the Lovell telescope at Jodrell Bank Observatory, or the Murriyang telescope at Parkes Observatory.

The MeerPipe pipeline produces ToAs from all observing epochs, even if the pulsar was not detected, or the observation was corrupted by interference.
Therefore some removal of bad data is needed, and ToAs were discarded if the pulsar was not detected, or if the ToA uncertainty was so great as to not provide useful information to the timing.
In most cases the lack of detection is due to pulsars with known nulling, or weak modes that are not detectable in the typical TPA timing observations with less than 2 minute durations.

\subsubsection{The timing model}
\label{timingmodel}
The final TPA timing data product was refined from the coarse timing model by means of applying a Bayesian noise modelling simultaneously with the timing model fitting.
The pulse noise model consists of a time-correlated achromatic spin noise, plus a time uncorrelated white noise component.
The achromatic spin noise is implemented using the Fourier-domain GP as described in \citet{temponest_physrevd} to model the noise by a power-law parameterised by dimensionless amplitude $A_\mathrm{red}$ and slope $\gamma$, and power spectral density given by
\begin{equation}
    P(f) = \frac{A_\mathrm{red}^2}{12\pi^2} \left(\frac{f}{f_\mathrm{yr}}\right)^{-\gamma} f_\mathrm{yr}^{-3},
\end{equation}
where $f_\mathrm{yr}$ is a frequency of 1 per year.
The white noise consists of the widely used \texttt{EFAC} and \texttt{EQUAD} parameters that scale and add in quadrature the independent white noise for each observation, plus an additional \texttt{ECORR} term correlated across observations at each epoch.
For ToAs with formal error given by $\sigma_i$, the white noise covariance matrix is given by
\begin{equation}
    \mathbf{C} = \mathrm{EFAC}^2\,\boldsymbol{\Sigma} + \mathrm{EQUAD}^2\,\boldsymbol{I} + \mathrm{ECORR}^2\,\boldsymbol{\Delta},
\end{equation}
where $\boldsymbol{\Sigma}$ is a diagonal matrix with $\Sigma_i = \sigma_i^2$, $\boldsymbol{I}$ is the identity matrix, and $\boldsymbol{\Delta}$ is a block-diagonal matrix such that $\Delta_{i,j} = 1$ for $i$, $j$ from the same observing epoch and $0$ otherwise (see e.g. \citealp{hv14} for a detailed description).

In addition to fitting for the noise parameters, we also marginalise over the following parameters in the pulsar ephemeris using the linearised model of \textsc{tempo2},
\begin{itemize}
    \item \texttt{F0}, \texttt{F1} and \texttt{F2} -- Spin frequency and two derivatives.
    \item \texttt{RAJ} and \texttt{DECJ} -- the position of the pulsar in J2000 coordinates.
    \item Glitch parameters (\texttt{GLF0}, \texttt{GLPH}), if required.
    \item Dispersion measure (\texttt{DM}) at each observing epoch.
    \item Arbitrary phase jumps for 6 of 8 sub-bands.
\end{itemize}
The arbitrary phase jumps between sub-bands absorb any evolution of the pulse profile with frequency.
In principle we may wish to fit for phase jumps between all 8 sub-bands, however, this would be fully covariant with the `zero phase' parameter that is always fit within tempo2.
Additionally, although the change in DM from epoch to epoch would be well measured, the average of the DM parameters would be unconstrained and hence not produce meaningful results.
Therefore, we must leave two of the sub-bands phase without a phase jump in order to define the long-term average DM, and so that the fit is not degenerate with the zero phase parameter.
In practice then the `zero phase' parameter becomes defined by the average phase of these two sub-bands, and the long-term average DM is defined by the phase difference between these two sub-bands.
In this analysis, we chose to keep fixed the top and bottom sub-bands, so that the average DM is determined using the most widely separated frequency channels, i.e. those centred at approximately 944 and 1623 MHz.
We note that since the phase jumps are constant in time, the choice of sub-bands to fit only affects the average DM, and not the epoch-to-epoch variability, which is the main focus of this work.

The noise modelling is performed using \textsc{run\_enterprise} \citep{kn23,run_enterprise}, which is based on the \textsc{enterprise} framework \citep{enterprise}, and sampling is performed using \textsc{emcee} \citep{emcee}.
The resulting noise model is then used to fit for the maximum likelihood values for the pulsar timing parameters using \textsc{tempo2}.

\subsection{Timeseries of the spin-down rate}
\label{nudot}
The timing residuals obtained for each pulsar are used to drive a time-domain GP, following Section 3.2 of \citet{brook16}. We use the same software and process used for the recent analysis of PSR~J0738$-$4042 by \citet{lower2023} to obtain $\dot{\nu}$ and its associated uncertainty. The GP is constructed with a single squared exponential kernel and a white noise term, and the second derivative of the GP can be directly computed to derive $\dot{\nu}(t)$.  
We sampled the GP hyperparameter posterior distributions using \textsc{bilby} \citep{ashton2019} as a wrapper for the \textsc{dynesty} nested sampling algorithm \citep{speagle2020}.
We then generated $\dot{\nu}$ points at each observing epoch directly from the GP. 
As always with this analysis, the assumption is that the residuals can be entirely explained by a time-variable $\dot{\nu}$.
In the majority of cases, the GP models the residuals extremely well, but there may be examples where the use of an additional kernel, together with strong priors on the hyperparameters, would result in a better fit. We also caution that the uncertainties for $\dot{\nu}$ can depend on the choice of kernel, and are only correct under the above assumptions.

\subsection{Dispersion Measure versus time}
\label{SecDMvsTime}

The DM of a pulsar measures the integral of the electron density, $n_e$ along the line of sight, $L$,
\begin{equation}
    {\rm DM} = \int_{0}^{L} n_e\!(s) \, \mathrm{d}s.
\end{equation}
Observationally, the DM can be obtained by measuring the time delay between the arrival times ($t_1$ and $t_2$) of the radio pulse at two different frequencies ($\nu_1$ and $\nu_2$) via
\begin{equation}
    {\rm DM} = K \frac{t_2 - t_1}{\nu_2^{-2} - \nu_1^{-2}}
\end{equation}
where $K$ is the dispersion constant and is equal to $2.410~\times~10^{-4}\,\textrm{MHz}^{-2}\,\textrm{cm}^{-3}\,\textrm{pc}\,\textrm{s}^{-1}$ (see \citealp{dm_constant} for a discussion of this value).
In practice the DM to a pulsar is not a constant, and we measure the DM for each epoch as part of the pulsar timing process described in Section \ref{timingmodel}.
It is worth noting that although the time dependence of the DM is usually well measured, there may be a systematic error in the DM if there is significant pulse profile evolution across the band, including effects from scattering. In order to get an indication of the scale of this systematic error, we compute the change in the estimated mean DM if we reference the DM against the highest or lowest pair of sub-bands rather than the most widely spaced sub-bands. It should be noted that this systematic error shifts all DM values up or down together, and does not change the perceived time variability.

\subsection{Rotation Measure versus time}
\label{SecRMvsTime}
The rotation measure (RM) of a pulsar measures the integral of the product of the electron density and the magnetic field parallel to the line of sight ($B_{||}$).
\begin{equation}
    {\rm RM} = \frac{e^3}{2\pi m_e^2c^4}\int_{o}^{L} n_e\!(s) B_{||}\!(s) \mathrm{d}s,
\end{equation}
where $m_e$ and $e$ are the mass and charge of an electron respectively.
Observationally, the RM can be obtained by measuring the change in the position angle (PA) of the linearly polarized radiation at two observing wavelengths ($\lambda_1$ and $\lambda_2$).
\begin{equation}
    {\rm RM} = \frac{{\rm PA}_2 - {\rm PA}_1}{\lambda_2^2 - \lambda_1^2}
\end{equation}
In practice, several different methods are employed to determine the RM and its error bar.

Here, the methodology as described in \citet{ijw19}, and implemented in \textsc{psrsalsa}\footnote{\url{https://github.com/weltevrede/psrsalsa}} \citep{wel16} is used. It is based on the RM synthesis technique (RMST; e.g. \citealt{bd05}), which optimises the degree of linear polarization as function of RM. 
Uncertainties are estimated via bootstrapping the data.
This relies on the identification of an on- and off-pulse region, which is done automatically using the same methodology as explained in \cite{song23}, and relies on the profile templates (see also Section~\ref{SectTemplatesAndToAs}).

The RM as measured at the telescope is determined for each epoch, resulting in a timeseries of RM measurements. Observations with very weak detections of the pulse profile, for instance because of RFI, were ignored in the analysis. This was done by visually inspecting pulse profiles of the in total 24,592 observations. To make this process more efficient, the pulse profiles of each pulsar are rank-ordered by the signal to noise ratio (S/N). The pulse profiles for each pulsar were visually inspected starting with the lowest S/N, and observations were rejected until a pulse profile of sufficient quality was encountered. Higher S/N detections were accepted without further visual inspection. In total 703 observations out of a total of 24,592 are excluded from the RM timeseries in this way.

The Earth's ionosphere contributes to the value of RM and this varies with time. This contribution was predicted using the \textsc{ionFR} package\footnote{\url{https://github.com/csobey/ionFR}} \citep{ssh+13}, which relies on global Total Electron Content (TEC) values\footnote{We used the TEC data products provided by JPL and distributed via\\ \url{https://cddis.nasa.gov/archive/gps/products/ionex/}.}. 
The predicted ionospheric contribution to the RM is subtracted in the figures (such as Fig.~\ref{fig:f1}, and the figures in the online material). In the tables of the online material the measured RM (which includes a contribution of the ionosphere) and the predicted RM contribution by the ionosphere are reported separately for each observation.

\subsection{Flux density}
\label{flux_density}
Flux calibration is carried out following the procedure described in \citet{census23}. In brief, the flux densities are scaled from the system counts via the radiometer equation and by using observations of pulsars at high galactic latitude (where the sky temperature is well understood) to calibrate the the sky temperatures of all other pulsars. The scatter in the observed flux density is generally much larger than the formal uncertainty.
Intrinsically, the flux density of a pulsar is observed to be largely stable with a low modulation index (see e.g. \citealt{kdj+21}). However, this is only true for observations of long duration (several thousand pulse rotations). In the monitoring program, a pulsar is typically observed for only 90~s or a few hundred rotations at best. The measured flux density will then fluctuate about some mean, depending on the random selection of single pulse flux densities from the underlying (and unknown) distribution. Furthermore, the traverse of the radio emission through the turbulent interstellar medium causes diffractive and refractive scintillation, which manifests itself as frequency and time dependent flux density variations. If these variations have comparable bandwidth and timescales to the observations, high modulation will be seen in the light curves \citep{kdj+21}. Finally, pulsars are observed to have a steep spectral index \citep{jvk+18,slm+22,census23} and averaging in frequency over a large bandwidth biases the flux densities towards lower frequencies than the nominal centre of the band.

With all these caveats in mind, here we simply average the pulse profile data in time and frequency for a given observation and present a single flux density for that observation. 

\begin{figure*}
\begin{center}
\begin{tabular}{cc}
\includegraphics[width=8cm,angle=0]{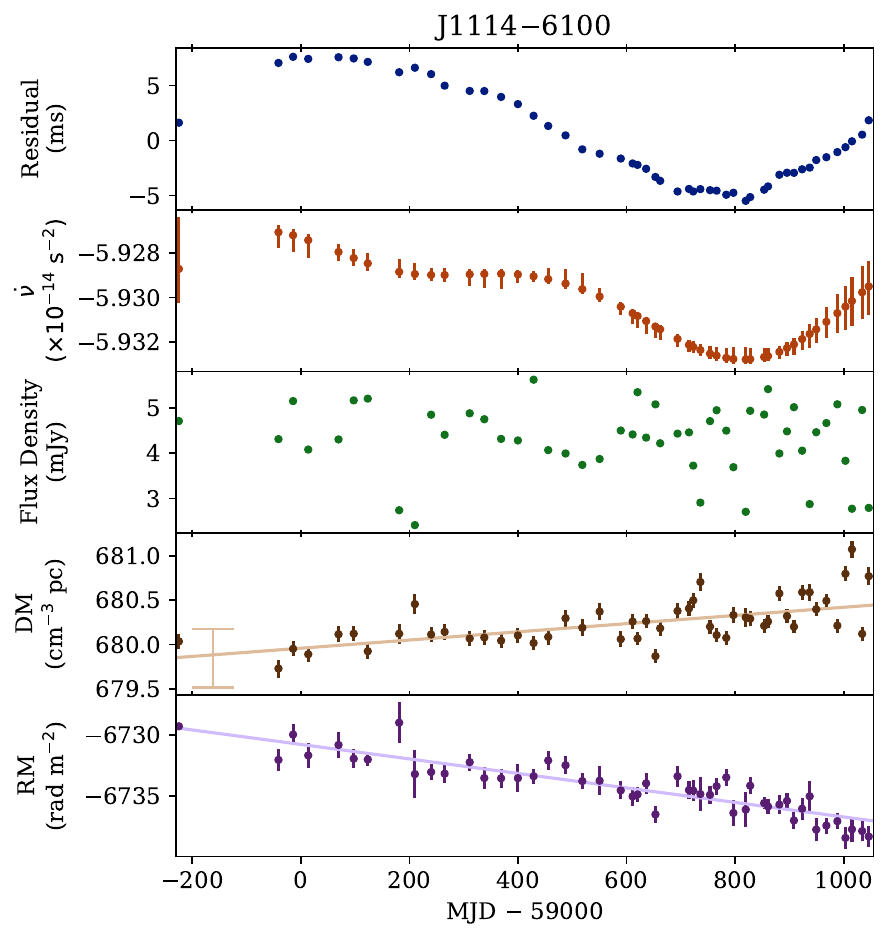} &
\includegraphics[width=8cm,angle=0]{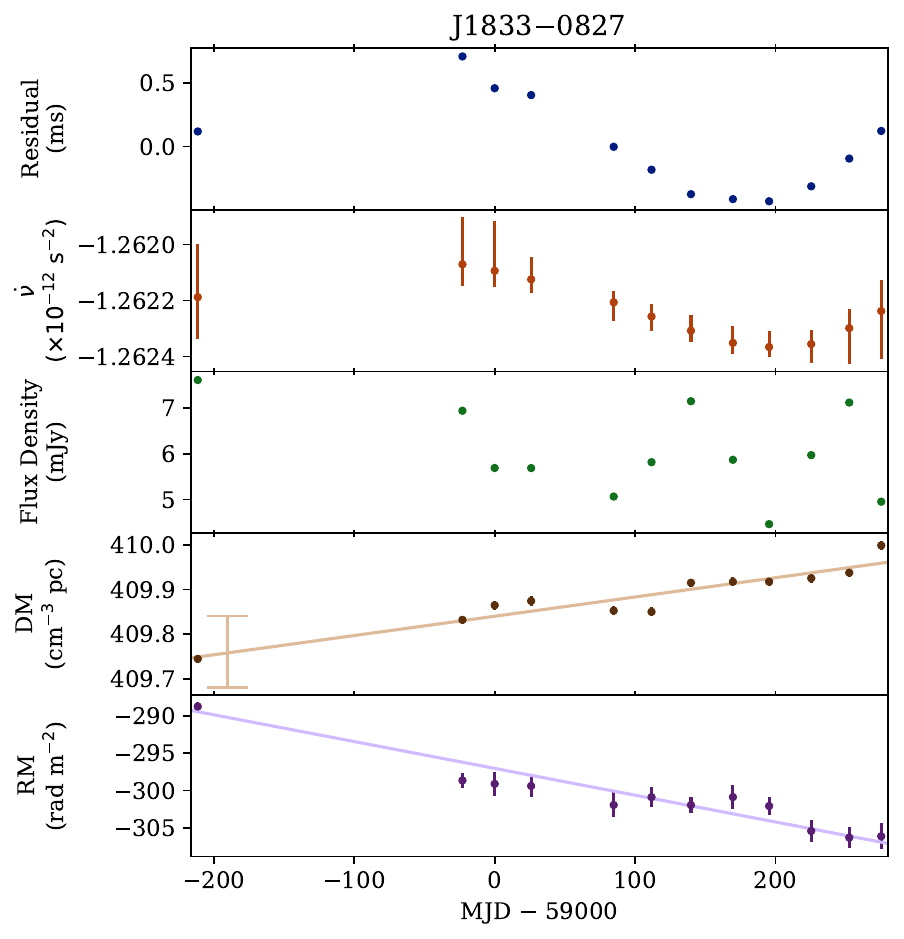} \\
\end{tabular}
\end{center}
\caption{As Figure~\ref{fig:bigplot}, but for PSRs~J1114$-$6100 (left) and J1833$-$0827 (right), two pulsars with significant RM and DM variations.}
\label{fig:bigrm}
\end{figure*}

\section{Description of tables and figures}
\label{tables_and_figures}
The data described in this section can be obtained from Zenodo, \url{doi.org/10.5281/zenodo.8430591}.
\subsection{Timing files}
Pulsar rotational ephemerides and ToAs are provided in \textsc{tempo2} compatible `par' and `tim' files. The ToAs in the tim files are annotated with a number of flags from the meerpipe reduction pipeline, including signal-to-noise ratio (\texttt{-snr}) and observing time span (\texttt{-length}). Each ToA is also associated with a relative pulse number to enable phase tracking in \textsc{tempo2}. Each pulsar is supplied with a pair of par files. The first par file contains the complete output of the noise modelling, as produced by the Bayesian pipeline. We also provide a simplified par file without the noise modelling and DM measurements specific to our data that is more easily transferred to other software and/or combined with data from other telescopes.

\subsection{Observation tables}
We provide a table containing the list of 597 pulsars we are monitoring, along with the MJD of the first and last observation and the number of data points. This table also includes the mean DM as well as the systematic offset when using either the two highest bands or the two lowest bands to define the DM. For each pulsar, in addition to the par and tim files described above, we also supply 6 ascii tables, one each for the residuals, dispersion measure, rotation measure, the rotation frequency derivative ($\dot{\nu}$) and flux density versus observation epoch. Each file has a one-line header with column labels that are described in Table \ref{coldesc}.
Measurements that are obviously spurious have been removed where, for instance, visual inspection of the observed profile suggested that the pulsar was not detected with a sufficient signal-to-noise ratio to make a measurement of a particular parameter.
Hence the number of rows in each table may differ due to the different sensitivity to each measured parameter.

\begin{table}
    \centering
    \caption{Description of the columns in the observation tables. Header lines give the file extension for each table. Epochs are topocentric values close to the midpoint of the observation.}
    \label{coldesc}
    \begin{tabular}{rl}

         Column Label & Description\\
         \hline
         \hline
         \multicolumn{2}{c}{\texttt{.restable}} \\
                  \hline

         MJD & Epoch of observation\\
         Res & Residual (s)\\
         Res\_err & Error on Residual (s) \\
         \hline
         \hline
         \multicolumn{2}{c}{\texttt{.dmtable}} \\
         \hline
          MJD & Epoch of observation\\
         DM & Dispersion Measure (cm$^{-3}$pc)\\
         DM\_err & Error on DM (cm$^{-3}$pc) \\
         \hline
         \hline
         \multicolumn{2}{c}{\texttt{.rmtable}} \\
         \hline
          MJD & Epoch of observation\\
         RM & Rotation Measure, as measured ($\mathrm{rad}\,\mathrm{m}^{-2}$) \\
         RM\_err & Error on RM ($\mathrm{rad}\,\mathrm{m}^{-2}$) \\
         RMiono & Ionospheric Rotation Measure ($\mathrm{rad}\,\mathrm{m}^{-2}$) \\
         RMiono\_err & Error on Ionospheric RM ($\mathrm{rad}\,\mathrm{m}^{-2}$) \\
           \hline
         \hline
         \multicolumn{2}{c}{\texttt{.f1table}} \\
         \hline
MJD & Epoch of observation\\
 F1 & GP predicted $\dot{\nu}$ (s$^{-2}$)\\
 F1\_upperr & 2-$\sigma$ upper error on $\dot{\nu}$ (s$^{-2}$)\\
 F1\_lowerr & 2-$\sigma$ lower error on $\dot{\nu}$ (s$^{-2}$)\\
\hline
         \hline
         \multicolumn{2}{c}{\texttt{.fluxtable}} \\
         \hline
                  MJD & Epoch of observation\\
         Flux & Phase Averaged Flux Density (mJy)\\
         Flux\_err & Error on Flux Density (mJy) \\
         \hline
    \end{tabular}
    
\end{table}

\subsection{Figures}
For each pulsar we provide a visual representation of the data described above. An example plot is given in Figure~\ref{fig:bigplot} for PSR~J1117$-$6154. The five panels show the timing residual (see Section~\ref{timing}), the $\dot{\nu}$ (see Section~\ref{nudot}), the DM (see Section~\ref{SecDMvsTime}), the RM (see Section~\ref{SecRMvsTime}) and the flux density (see Section~\ref{flux_density}) as a function of MJD. 1-$\sigma$ errors are shown on all values.

\section{Data highlights}
\label{highlights}

\subsection{The interstellar medium}
The strength of turbulence in the ISM can be characterised by a power-law over many orders of magnitude bounded by the inner and outer scales \citep{rcb84}. The various scales can be explored by scintillation observations (both diffractive and refractive), DM variations with time and RM variations with time. Given dense samples and a long time-baseline, the structure function is the implement of choice for exploring the nature of the turbulence (e.g. \citealp{cr98}).
However, the structure function is the average power as a function of lag, and requires being able to derive multiple independent samples of the process (at the lag of interest) from the data.
For most pulsars in our sample, we are only sensitive to power on the timescale of our dataset, and hence the structure functions contains little meaningful data. 
We therefore resort to simply measuring any gradient in the DM versus time data, which gives an indication of the strength of the DM variations on the longest timescales. 
Table~\ref{tab:sigdm} records those pulsars for which we measure a gradient greater than 3-$\sigma$ from zero. There are 87 pulsars with a significant gradient (out of 597), a significantly larger fraction than the 4 out of 160 pulsars reported in \citet{pkj+13}, highlighting the improved sensitivity in the TPA data.

One also expects that the RM should change with time, particularly if the DM is also changing. In addition, however, changes in the average magnetic field strength along the line of sight should also be observed, particularly if the pulsar is in or behind a magnetic structure such as a supernova remnant. In Table~\ref{tab:sigrm} we list those pulsars for which the slope of RM versus time is more than 3-$\sigma$ significant. There are 58 such pulsars of which 15 are in common with pulsars with a DM gradient.

There are two pulsars which show very large changes in both DM and RM as shown in Figure~\ref{fig:bigrm}. PSR~J1114$-$6100 has a very large RM for its location and was discussed extensively in \citet{jsd+21}. Over the course of the monitoring campaign its RM has changed by some 6~rad$\,$m$^{-2}$ and its DM by 1~cm$^{-3}$pc. PSR~J1833$-$0827 had a large negative DM slope in \citet{pkj+13}. Here we see that the DM slope is positive and that the RM has changed by 15~rad$\,$m$^{-2}$ in less than 2~yr. As \citet{pkj+13} pointed out, this pulsar has an associated X-ray pulsar wind nebula \citep{eit+11}, making the nebula the most likely source of the fluctuations in RM and DM.

\subsubsection{DM slopes vs DM}
If the DM slopes we observe in our sample originate in discrete structures in the ionised ISM (IISM) then we may expect that the amplitude would scale with the square-root of the number of such structures encountered, and hence roughly with the square-root of the DM.
Alternatively, we may expect that the observed DM slopes come from power on the longest length scales in the turbulent IISM, which are expected to be at least 100~pc~\citep{armstrong81}, far longer than the scales probed by our $\sim 4$ year dataset.
To get a feeling for how the DM variations due to IISM turbulence may be expected to scale with DM, we can look at scattering, which is another effect of IISM turbulence.
The `level of turbulence' in the IISM can be estimated from measurements of interstellar scattering \citep{cordes86}, where the scattering measure (SM) is the path-integrated variance in the electron density along the line of sight \citep{bhvf93}.
The SM can be estimated from observations of scattering timescale, $\tau_\mathrm{scatt}$, where for Kolmogorov turbulence, 
\begin{equation}
\label{SM}
\mathrm{SM} \propto (\tau_\mathrm{scatt}/D)^{5/6},
\end{equation}
where $D$ is the distance to the pulsar \citep{cwf+91}.
If the power in the IISM is independent of location then one expects to find $\tau_\mathrm{scatt} \propto \mathrm{DM}^{2.2}$, however measurements of $\tau_\mathrm{scatt}$ show evidence for inhomogenities in the level of turbulence, with \citet{kmn+15} finding
\begin{equation}
\label{tscat}
\tau_\mathrm{scatt} \propto \mathrm{DM}^{2.2}\left(1+1.94\times10^{-3}\mathrm{DM}^{2}\right)
\end{equation}
over a the full range of $\mathrm{DM}$ values observed in the pulsar population.

Whilst the SM is a measure of the variance of the electron density, the amplitude of the DM variations should scale with the standard deviation of the DM, and hence all else being equal, the typical amplitude of DM slope to a pulsar should scale with the square-root of the expected SM \citep{bhvf93}.  Hence, we can insert Equation \ref{tscat} into Equation \ref{SM} and square-root. Making also the simplifying assumption that $D\propto \mathrm{DM}$ we can get a scaling relation
\begin{equation}
    |\mathrm{DM\, slope}| \propto \mathrm{DM}^{1/2}\left(1+1.94\times10^{-3}\mathrm{DM}^{2}\right)^{5/12}.
    \label{dm1_function_of_dm}
\end{equation}

Hence we may expect that the exponent of DM slope to evolve slightly with DM, from $1/2$ at low DM, to $4/3$ at high DM, but in practice the intrinsic scatter means that this does not seem distinguishable from a DM slope linearly proportional to DM over the TPA sample.

\begin{figure}
\begin{center}
\includegraphics[width=8.5cm,angle=0]{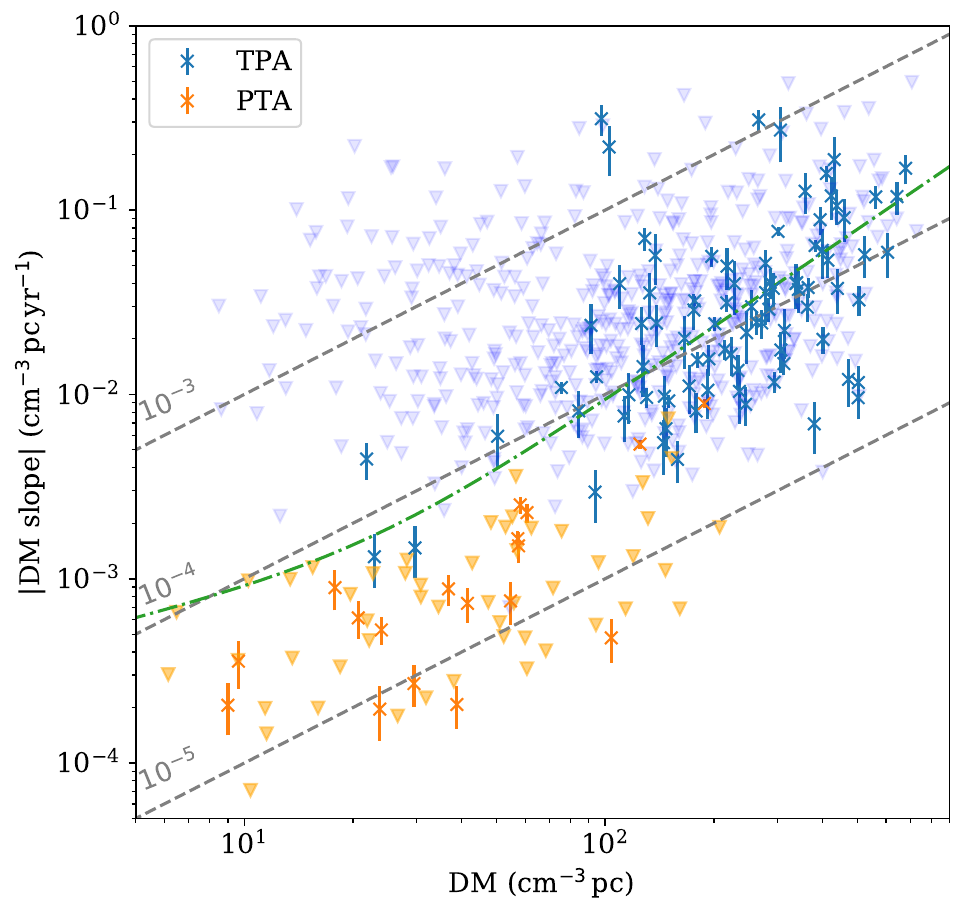}
\end{center}
\caption{DM slope as a function of DM for the MeerTime TPA pulsars, and the MSPs in the MeerTime PTA data release. Downward triangles mark 2-$\sigma$ upper limits. Diagonal dashed lines of proportionality are shown for reference. The dot dashed curved line shows the proportionality relationship derived from scattering measurements from Equation \ref{dm1_function_of_dm}.}
\label{fig:dm1_vs_dm}
\end{figure}

Indeed, we find that the measured DM gradient for the TPA pulsars appears consistent with linear increase with DM with a slope of the order of $10^{-4}\times \mathrm{DM}$ per year.
Figure \ref{fig:dm1_vs_dm} shows this for the TPA pulsars, and also for the sample of MSPs from the MeerTime PTA data release \citep{miles2023} which is much more sensitive to DM variations due to the nature of precision MSP timing.
Although both seem to scale with DM, the TPA sample shows a factor of around 5 times larger DM slopes for a given DM.
PTA pulsars are, due to selection effects in the discovery of pulsars, more likely to be at lower DM than the TPA sample, and hence some of the observed difference may be due to the inhomogenties in the IISM such as those which give rise to Equation \ref{dm1_function_of_dm}.
However, as seen in the dot-dashed line in Figure \ref{fig:dm1_vs_dm}, the increased DM variation at large DM does not seem strong enough to explain the difference.

\begin{figure}
\begin{center}
\includegraphics[width=8.5cm,angle=0]{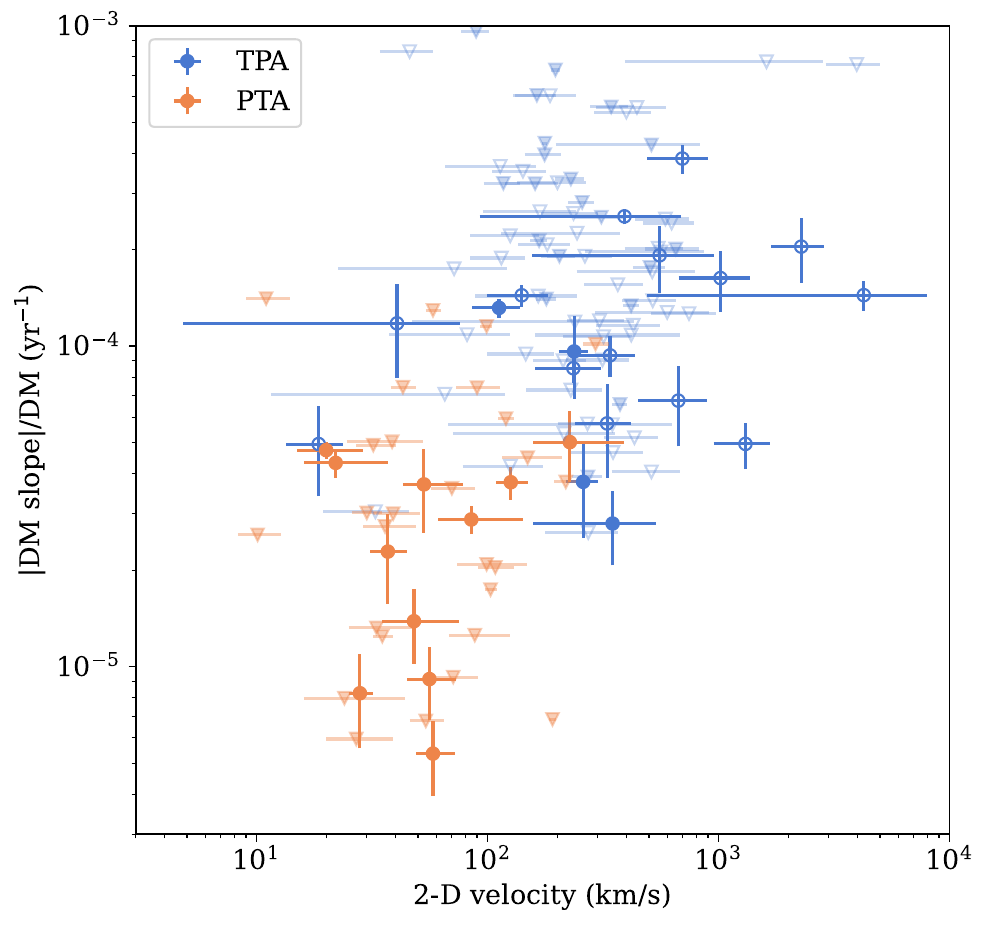}
\end{center}
\caption{DM slope, scaled by DM, as a function of velocity. For TPA pulsars velocity is inferred from proper motion measurements and distance estimates taken from \textsc{psrcat}. PTA velocities taken from \citet{sbf+24}. Not all pulsars have proper motions measured. Downward triangles mark 2-$\sigma$ upper limits. Pulsars with parallax derived distances are shown with filled symbols, those with DM-derived distances are shown with open symbols.}
\label{fig:dm1_vs_vel}
\end{figure}

An important consideration for interpreting these results is that the rate at which we observe DM fluctuations does not only depend on the scale of the turbulence in the IISM, but also on the rate at which the line of sight to the pulsar traverses the IISM.
For most pulsars, the line of sight velocity is dominated by the velocity of the pulsar, therefore, we may suppose that the the difference in scale of DM gradients may instead be due to the  smaller velocity distribution observed in MSPs compared to the general pulsar population \citep{hobbs05,sbf+24}.
Figure \ref{fig:dm1_vs_vel} shows the fractional DM slope as a function of an estimate of the pulsar 2-D velocity.
For the PTA pulsar sample, the velocities are taken from \citep{sbf+24}, which are based on VLBI or timing parallax and proper motions.
This velocity for TPA pulsars is estimated from proper motion and distance as recorded in \textsc{psrcat}. Parallax measurements are used if available, or otherwise distance is inferred from the DM using the `YMW16' electron density model \citep{ymw17}.
It should be noted that the uncertainties in the velocity of TPA pulsars are not rigorous as distance estimates can be unreliable, and we have made an assumption of normally distributed independent uncertainties on all measurements.
We refer readers to \citet{vic17} or \citet{sbf+24} for a complete and careful study of the intrinsic distribution of pulsar velocities.
Nevertheless, studies of pulsar velocities seem clear that the `normal' pulsar population sampled by the TPA has an average velocity at least 3 times greater than the PTA sample, and there is evidence that a large fraction of the normal pulsars have velocities 5-10 times greater \citep{vic17,sbf+24}.
Therefore we feel that that the larger DM variations observed in the TPA pulsars can be reasonably attributed to the fact that the line of sight to the pulsar traverses the ISM at a rate up to an order of magnitude faster than for the PTA sample.
If one assumes that the observed DM slope scales linearly with pulsar velocity, and that the dependence on DM is also close to linear over the typical pulsar distance, we suggest a `rule of thumb' for predicting the magnitude of DM variations in a pulsar by
\begin{equation}
|\mathrm{DM slope}| \sim 0.01\, \mathrm{cm^{-3}\,pc\,yr^{-1}} \left(\frac{\mathrm{DM}}{100\,\mathrm{cm^{-3}\,pc}}\right) \left(\frac{v}{300\,\mathrm{km\,s^{-1}}}\right),  
\end{equation}
where $v$ is the pulsar's velocity. We note that in reality there are large, and potentially systematic variations in this value due to the complex structure of the IISM.


\subsubsection{RM slopes vs DM}
\begin{figure}
\begin{center}
\includegraphics[width=8.5cm,angle=0]{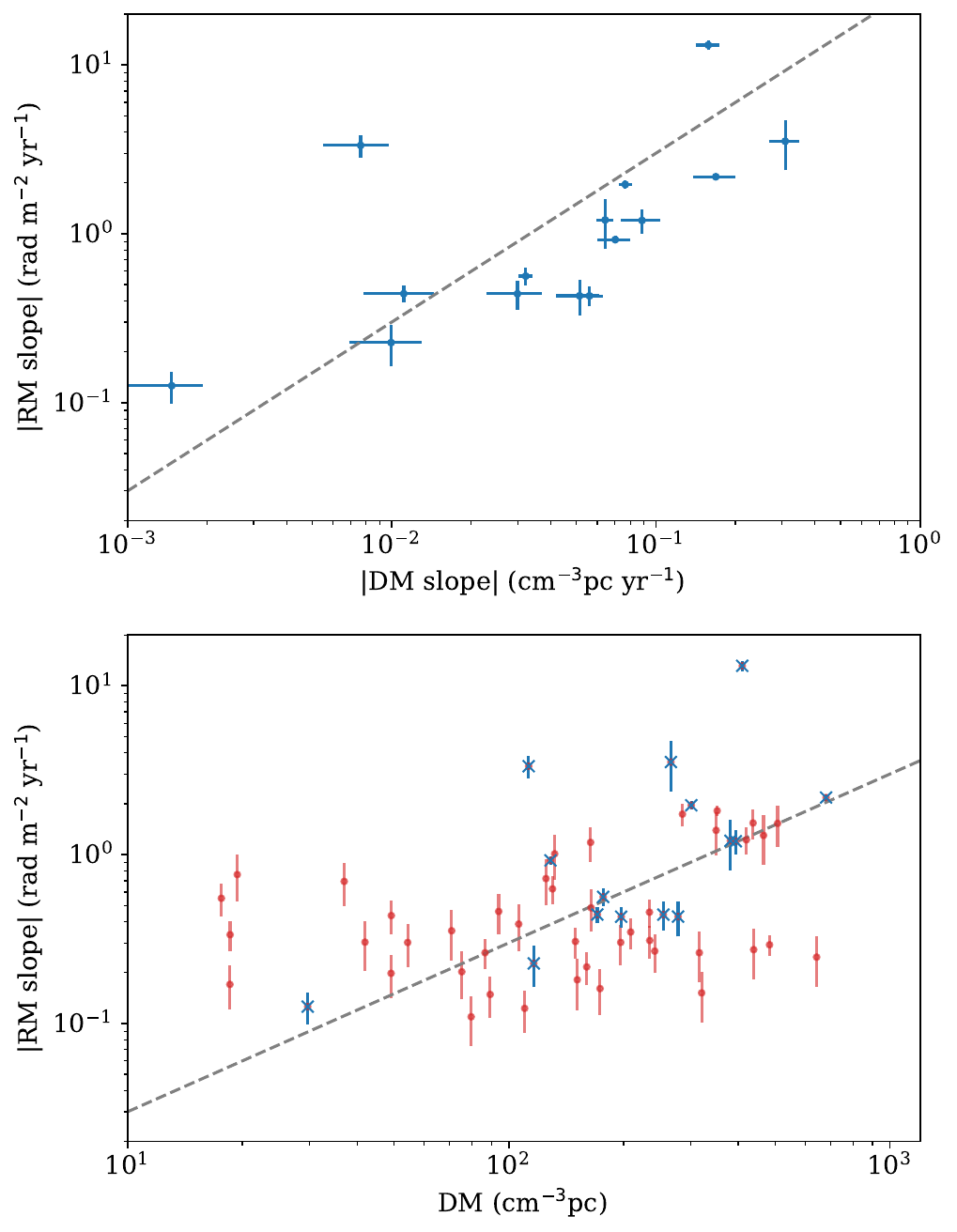}
\end{center}
\caption{Upper panel: Magnitude of RM slope vs DM slope for the 15 pulsars where we measure both RM and DM slopes. Lower panel: Magnitude of RM slope vs DM for all pulsars where RM slopes were detected (red points) and those with both RM and DM slopes (blue crosses). In both panels the black gray dashed line is a line of linear proportionality.}
\label{fig:rm_vs_dm}
\end{figure}

There are 15 pulsars for which we measure a significant DM slope and a significant RM slope. These are plotted in the upper panel of Figure \ref{fig:rm_vs_dm}.
There is a clear correlation between the magnitude of the RM and DM slopes in these pulsars.
There does not appear to be any correlation between the sign of the DM slope and the relative sign of the RM slope (i.e. if the magnitude of RM is increasing or decreasing) amongst these pulsars.
Put together, this suggests that for these pulsars at least, the DM slopes and RM slopes are likely arising from electron density variations in the same structures in the IISM, though these structures do not dominate the overall RM.

The lower panel of Figure \ref{fig:rm_vs_dm} also shows that more broadly across the TPA sample, the amplitude of the RM gradients do not scale strongly with DM, perhaps closer to $\sqrt{\mathrm{DM}}$ or even flat, rather than the linear growth seen in the DM gradients.
A scaling with $\sqrt{\mathrm{DM}}$ may suggest that the RM slopes originate from a sum over discrete structures in the ISM, perhaps dominated by variations in the local magnetic fields which would be transparent to the DM.
A sum over discrete structures could be expected to increase as the square root of the number of structures on the line of sight, and hence roughly as $\sqrt{\mathrm{DM}}$ \citep{bhvf93}.
Hence, the observed RM slopes may have two origins, those that arise from the same turbulent IISM effects that cause the DM slopes, and hence scale proportionally with the DM slopes, and those that are caused by discrete magnetised structures on the line of sight and hence scale most closely to the square-root of the distance to the pulsar.

\subsection{Glitches}
Table~\ref{tab:glitch} shows the preliminary results for the glitches in the sample as the fractional change in the spin frequency $\Delta\nu / \nu$ and its error. There are 11 glitches in 9 pulsars. Further studies are ongoing to fully characterise these glitches in terms of transient events or changes in $\dot{\nu}$.

Glitches of comparable size to those listed here have previously been seen in 7 of the pulsars (see compilations in \citealt{ljd+21} and \citealt{bsa+22}). For two of the pulsars we have detected glitches for the first time. PSR~J1341$-$6023 is a canonical middle-aged pulsar; these pulsars undergo large but rare glitches. In contrast, PSR~J1935+2025 has spin parameters similar to the `Vela-like' pulsars, which undergo quasi-periodic glitches on timescales of years to decades. In our three years of monitoring this pulsar we have detected two glitches, one relatively small in amplitude and the other relatively large. Again, this behaviour is common to the Vela-like pulsars.
\begin{table}
\caption{Preliminary parameters for glitches for the pulsars in the sample. Note that there is only one ToA prior to the glitch in PSR J1826$-$1334, hence the glitch parameters are estimated by reference to pre-glitch observations from Jodrell Bank Observatory.}
\label{tab:glitch}
\begin{center}
\begin{tabular}{crrl}
\hline
\hline
JNAME & Epoch & \multicolumn{2}{c}{$\Delta\nu / \nu$} \\
& (MJD) & \multicolumn{2}{c}{$(\times 10^{-9})$} \\
\hline
J1019$-$5749 & 59182 & $8.3$ & $\pm1.0$ \\
J1019$-$5749 & 59265 & $398.3$ & $\pm2.5$ \\
J1341$-$6023 & 59450 & $5774.98$ & $\pm0.24$ \\
J1453$-$6413 & 59015 & $1.14$ & $\pm0.13$ \\
J1524$-$5625 & 59297 & $2076$ & $\pm6$ \\
J1803$-$2137 & 58920 & $4702$ & $\pm11$ \\
J1826$-$1334 & 58879 & $2470$ & $\pm10$ \\
J1836$-$1008 & 58950 & $32.8$ & $\pm0.9$ \\
J1837$-$0604 & 59610 & $3290$ & $\pm50$ \\
J1935+2025 & 59178 & $93$ & $\pm6$ \\
J1935+2025 & 59591 & $2840$ & $\pm40$ \\

\hline
\end{tabular}
\end{center}
\end{table}

\subsection{Variations in $\dot{\nu}$}

The observed $\dot{\nu}$ variations can be broadly classified into 5 classes: {(i)} timeseries that show a constant slope in $\dot{\nu}$, i.e. where there is a significant $\ddot{\nu}$; {(ii)} constant timeseries with no $\dot{\nu}$ variations; {(iii)} timeseries characterized by smooth, long timescale variations; {(iv)} timeseries that show significant short term variability; and {(v)} timeseries that show a periodic or quasi-periodic signature.

\begin{figure*}
\begin{center}
\begin{tabular}{cc}
\includegraphics[width=8.5cm,angle=0]{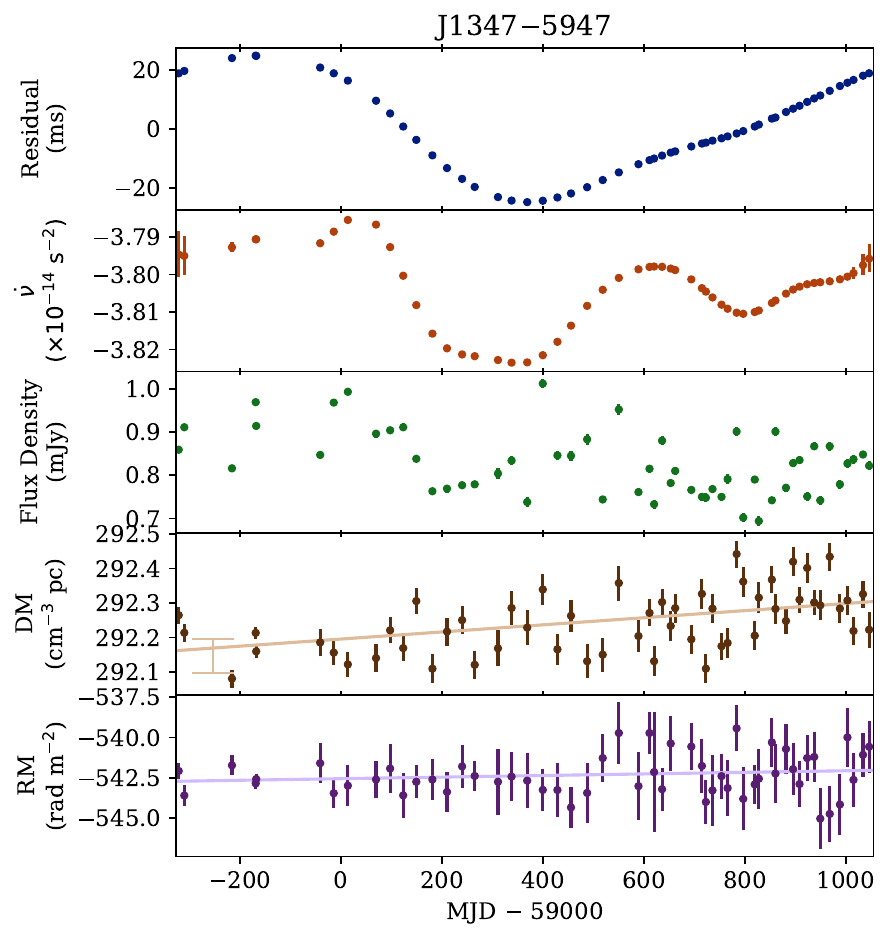} &
\includegraphics[width=8.5cm,angle=0]{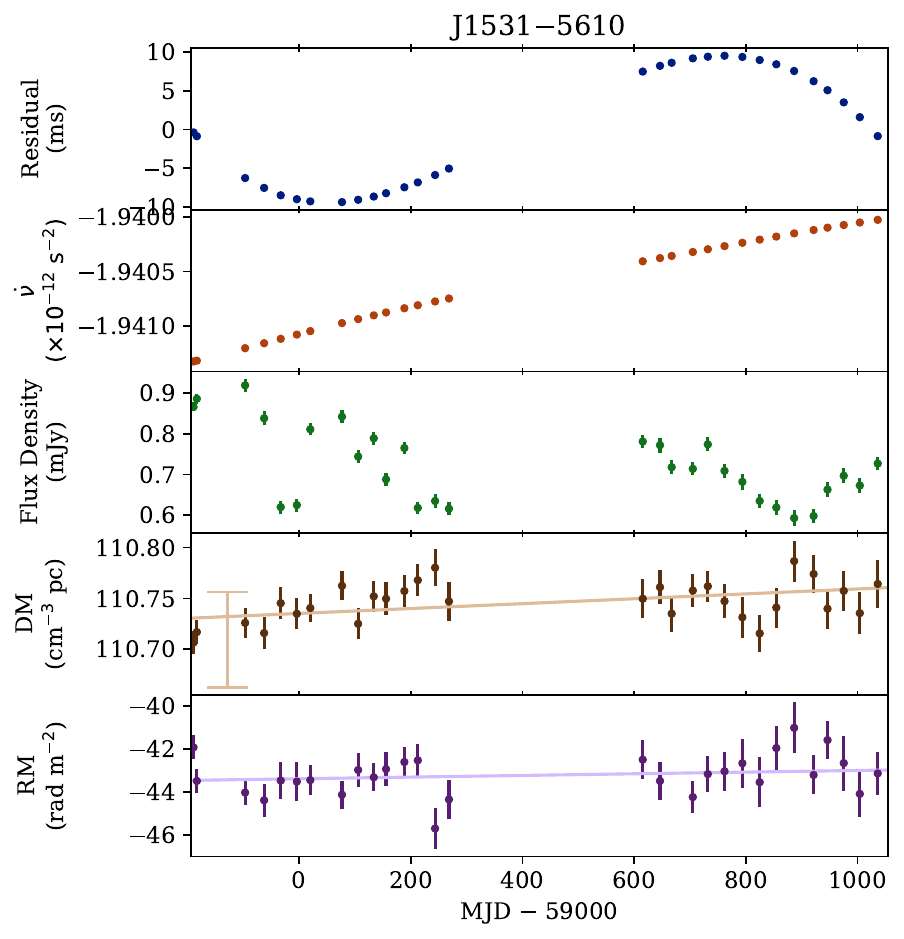} \\
\includegraphics[width=8.5cm,angle=0]{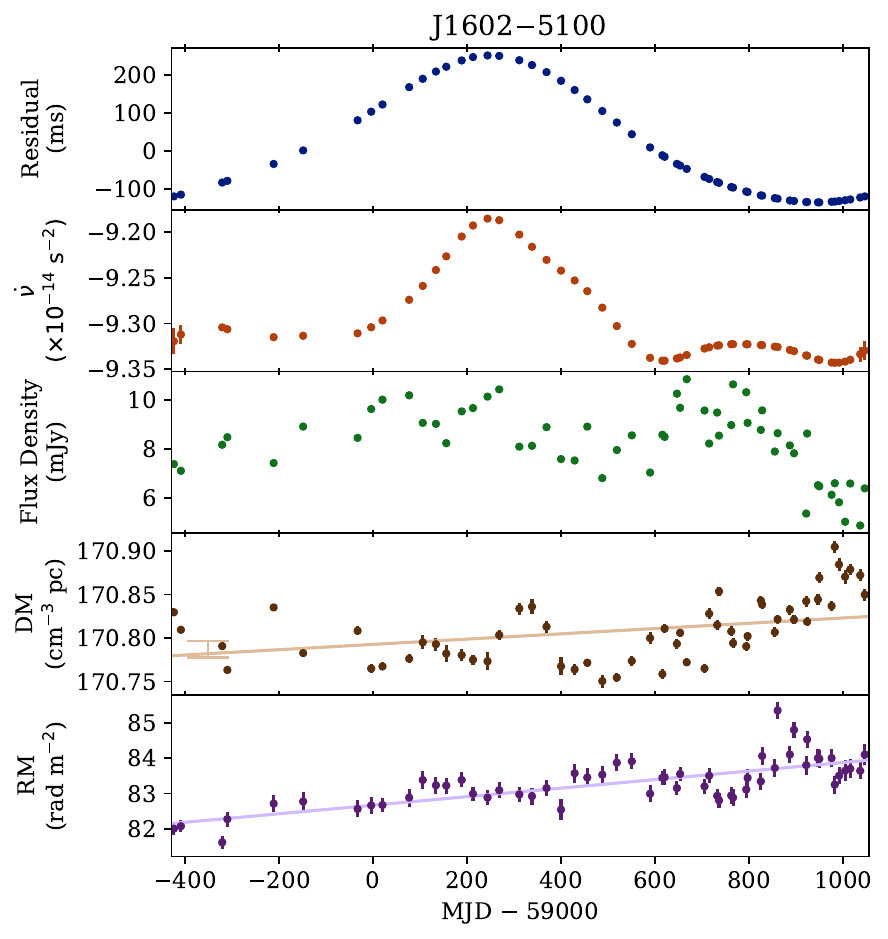} &
\includegraphics[width=8.5cm,angle=0]{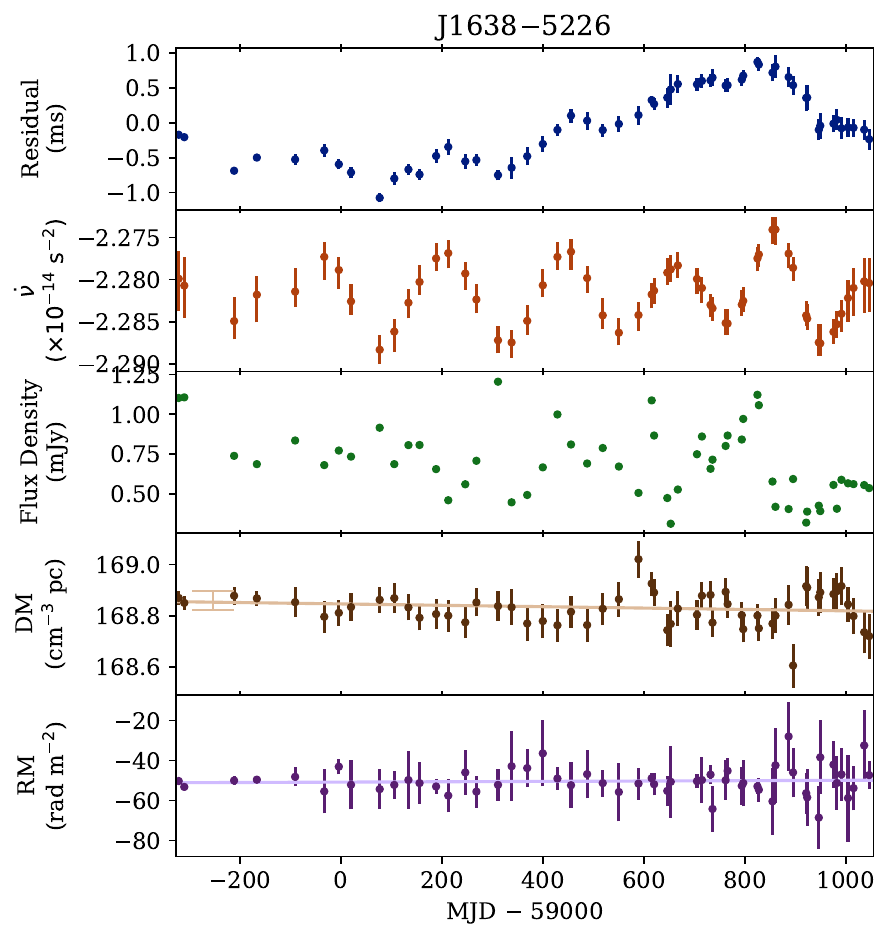} \\
\end{tabular}
\end{center}
\caption{As Figure~\ref{fig:bigplot}, but for PSRs~J1347$-$5947 (top left), J1531$-$5610 (top right), J1602-5100 (bottom left) and J1638$-$5226 (bottom right), pulsars with notable $\dot{\nu}$ variations.}
\label{fig:f1}
\end{figure*}

Four examples of $\dot{\nu}(t)$ behaviour are shown in Figure~\ref{fig:f1}.
PSR~J1347$-$5947 shows a low-level stochastic wandering of $\dot{\nu}$, seemingly typical behaviour for this type of moderate $\dot{E}$ pulsars. PSR~J1531$-$5610 shows a straight line in $\dot{\nu}$, indicative of a large value of $\ddot{\nu}$. Indeed, $\ddot{\nu}$ was measured for this pulsar by \citet{pjs+20} who obtained a value of $1.37(2)\times 10^{-23}$~s$^{-3}$. The value we measure here, $1.23(5)\times 10^{-23}$~s$^{-3}$ over a much shorter time-span is similar to their value. PSR~J1602$-$5100 was shown by \citet{brook16} to undergo a profile change accompanied by a step change in $\dot{\nu}$ around MJD 54700. In the MeerKAT data we see a similar $\dot{\nu}$ event around MJD 59200 which also lasts for several hundred days. Unlike the event in \citet{brook16}, here the $\dot{\nu}$ returns to its original pre-event value. PSR~J1638$-$5226 is an example of a pulsar with highly regular quasi-periodic oscillations in $\dot{\nu}$ with a time-scale of some 220~days. Such quasi-periodic behaviour is seen widely across the pulsar population \citep{hlk10,nitu22} and thought to be linked to emission changes \citep{lhk+10}.

\section{Conclusions}
This paper presents the first data release from the MeerTime TPA, and we highlight a few of the most striking features from the 597 pulsars.
A range of spin-down behaviours is observed, and future work will attempt to identify correlations with profile shape changes (or other properties).
We also observe significant DM and RM variations in 15\% and 13\% of the pulsars over the 4-year timescale of our observations.
These DM variations seem larger than a similar sample of MSPs, and we attribute this mainly to the different velocity dispersion in these two families of pulsars, as the line of sight traverses the ISM more quickly in the normal pulsar population.
A full analysis of all the data in this data release is outside the scope of this data release paper, but we anticipate that the data will be used for a wide range of projects within the TPA and externally.

\section*{Acknowledgements}
The MeerKAT telescope is operated by the South African Radio Astronomy Observatory (SARAO), which is a facility of the National Research Foundation, an agency of the Department of Science and Innovation. 
SARAO acknowledges the ongoing advice and calibration of GPS systems by the National Metrology Institute of South Africa (NMISA) and the time space reference systems department department of the Paris Observatory. 
PTUSE was developed with support from the Australian SKA Office and Swinburne University of Technology. 
This work made use of the OzSTAR national HPC facility at Swinburne University of Technology.
MeerTime data is housed on the OzSTAR supercomputer.
The OzSTAR program receives funding in part from the Astronomy National Collaborative Research Infrastructure Strategy (NCRIS) allocation provided by the Australian Government.
Pulsar research at Jodrell Bank Centre for Astrophysics is supported by an STFC Consolidated Grant (ST/T000414/1; ST/X001229/1). Parts of this research were conducted by the Australian Research Council Centre of Excellence for Gravitational Wave Discovery (OzGrav), through project number CE170100004.
%%%%%%%%%%%%%%%%%%%%%%%%%%%%%%%%%%%%%%%%%%%%%%%%%%
\section*{Data Availability}
The data tables and figures in the TPA data release are available at \url{doi.org/10.5281/zenodo.8430591}.
Profile data, ToAs and other automatically produced data products can be obtained from the MeerTime Portal (\url{https://pulsars.org.au}).
We request that authors cite this paper and include the zenodo DOI when making use of the TPA data release in future publications.



\bibliographystyle{mnras}
\bibliography{tpa_timing} % if your bibtex file is called example.bib

%\begin{thebibliography}{99}
%\end{thebibliography}

%%%%%%%%%%%%%%%%%%%%%%%%%%%%%%%%%%%%%%%%%%%%%%%%%%

%%%%%%%%%%%%%%%%% APPENDICES %%%%%%%%%%%%%%%%%%%%%
\appendix
\section{Tables of RM and DM variations}
\include{sigdm}

\include{sigrm}
\bsp	% typesetting comment
\label{lastpage}
\end{document}

%% file: sigdm.tex
\begin{table*}
\caption{Pulsars for which the measured slope in DM is more than 3-$\sigma$ away from zero. Reduced $\chi^2$ of the linear fit is also given as an indication of the goodness of fit.}
\label{tab:sigdm}
\begin{center}
\begin{tabular}{crll|crll|crll}
\hline
\hline
PSR & \multicolumn{2}{c}{DM slope} & $\chi^2$ & PSR & \multicolumn{2}{c}{DM slope} & $\chi^2$ & PSR & \multicolumn{2}{c}{DM slope} & $\chi^2$\\
 & \multicolumn{2}{c}{($10^{-3} \mathrm{cm}^{-3}\mathrm{pc}\,\mathrm{yr}^{-1}$)}  & & & \multicolumn{2}{c}{($10^{-3} \mathrm{cm}^{-3}\mathrm{pc}\,\mathrm{yr}^{-1}$)} & & & \multicolumn{2}{c}{($10^{-3} \mathrm{cm}^{-3}\mathrm{pc}\,\mathrm{yr}^{-1}$)} &\\
\hline
J0134$-$2937 & $-4.5$ & $\pm1.0$ & 0.22 & J0807$-$5421 & $20.1$ & $\pm6.4$ & 1.3 & J0849$-$6322 & $23.7$ & $\pm7.2$ & 0.96 \\
J0904$-$4246 & $-9.8$ & $\pm2.7$ & 1.0 & J0908$-$4913 & $-15.4$ & $\pm1.5$ & 97 & J1013$-$5934 & $6.9$ & $\pm2.2$ & 3.2 \\
J1015$-$5719 & $51.4$ & $\pm9.5$ & 1.0 & J1016$-$5857 & $-89$ & $\pm15$ & 1.2 & J1019$-$5749 & $486$ & $\pm70$ & 0.23 \\
J1042$-$5521 & $-17.4$ & $\pm4.4$ & 2.9 & J1047$-$6709 & $9.9$ & $\pm3.1$ & 15 & J1048$-$5832 & $-70$ & $\pm10$ & 13 \\
J1057$-$5226 & $1.47$ & $\pm0.46$ & 2.0 & J1105$-$6107 & $24.4$ & $\pm4.4$ & 14 & J1114$-$6100 & $169$ & $\pm31$ & 5.1 \\
J1123$-$4844 & $2.95$ & $\pm0.94$ & 1.5 & J1123$-$6259 & $16.5$ & $\pm4.0$ & 0.36 & J1306$-$6617 & $-105$ & $\pm23$ & 1.1 \\
J1326$-$6408 & $9.6$ & $\pm2.3$ & 1.6 & J1341$-$6023 & $-38.0$ & $\pm4.7$ & 2.2 & J1347$-$5947 & $37.6$ & $\pm7.8$ & 3.8 \\
J1359$-$6038 & $11.7$ & $\pm1.5$ & 44 & J1412$-$6111 & $22.2$ & $\pm6.7$ & 0.65 & J1507$-$6640 & $-9.6$ & $\pm1.2$ & 2.1 \\
J1534$-$4428 & $57$ & $\pm18$ & 0.55 & J1539$-$5626 & $28.7$ & $\pm6.3$ & 0.50 & J1546$-$5302 & $-40.5$ & $\pm7.5$ & 0.91 \\
J1550$-$5242 & $-40$ & $\pm11$ & 0.75 & J1600$-$5044 & $-24.6$ & $\pm3.6$ & 25 & J1602$-$5100 & $11.1$ & $\pm3.3$ & 31 \\
J1609$-$4616 & $9.22$ & $\pm0.74$ & 0.62 & J1611$-$5209 & $-14.1$ & $\pm4.4$ & 1.3 & J1632$-$4621 & $118$ & $\pm16$ & 2.2 \\
J1633$-$5015 & $-60$ & $\pm18$ & 4.0 & J1637$-$4553 & $-15.6$ & $\pm3.0$ & 1.4 & J1649$-$3805 & $17.5$ & $\pm2.1$ & 0.44 \\
J1656$-$3621 & $-40$ & $\pm13$ & 1.2 & J1658$-$4958 & $-10.5$ & $\pm3.1$ & 0.84 & J1705$-$1906 & $1.32$ & $\pm0.43$ & 0.88 \\
J1709$-$4429 & $10.89$ & $\pm0.85$ & 3.1 & J1722$-$3207 & $24.2$ & $\pm5.7$ & 1.3 & J1723$-$3659 & $-29.9$ & $\pm7.1$ & 1.1 \\
J1739$-$2903 & $24.3$ & $\pm6.3$ & 2.3 & J1744$-$5337 & $-40$ & $\pm11$ & 0.49 & J1752$-$2806 & $5.9$ & $\pm1.9$ & 20 \\
J1759$-$2205 & $-32.1$ & $\pm2.0$ & 20 & J1807$-$2715 & $-14.7$ & $\pm2.0$ & 2.0 & J1808$-$3249 & $-6.5$ & $\pm1.9$ & 1.6 \\
J1809$-$1917 & $-56.0$ & $\pm7.2$ & 1.6 & J1809$-$2109 & $-64.2$ & $\pm4.7$ & 1.6 & J1813$-$2113 & $91$ & $\pm24$ & 1.0 \\
J1816$-$1729 & $-57$ & $\pm15$ & 1.1 & J1820$-$0427 & $-8.1$ & $\pm2.3$ & 8.2 & J1828$-$0611 & $-29.7$ & $\pm5.2$ & 1.2 \\
J1828$-$1101 & $59$ & $\pm16$ & 1.3 & J1829$-$1751 & $-31.3$ & $\pm3.3$ & 30 & J1830$-$0131 & $313$ & $\pm60$ & 0.24 \\
J1831$-$0823 & $-21.6$ & $\pm6.9$ & 1.3 & J1832$-$0827 & $76.6$ & $\pm4.3$ & 0.68 & J1833$-$0827 & $158$ & $\pm16$ & 5.1 \\
J1834$-$1202 & $-37.1$ & $\pm9.6$ & 0.51 & J1835$-$1106 & $35.5$ & $\pm9.8$ & 2.5 & J1842$+$1332 & $-220$ & $\pm67$ & 0.75 \\
J1843$-$0806 & $50$ & $\pm13$ & 1.6 & J1845$-$0635 & $-53$ & $\pm11$ & 1.4 & J1852$-$0127 & $-188$ & $\pm62$ & 0.96 \\
J1857$+$0212 & $32.6$ & $\pm6.1$ & 1.9 & J1901$+$0331 & $19.9$ & $\pm3.3$ & 5.3 & J1902$+$0615 & $11.6$ & $\pm2.7$ & 2.4 \\
J1903$+$0135 & $8.8$ & $\pm2.1$ & 9.0 & J1904$+$0004 & $-13.5$ & $\pm2.8$ & 2.7 & J1904$+$0738 & $36$ & $\pm12$ & 0.75 \\
J1904$+$0800 & $-38$ & $\pm10$ & 0.59 & J1906$+$0641 & $12.0$ & $\pm3.5$ & 0.72 & J1906$+$0912 & $309$ & $\pm41$ & 0.50 \\
J1908$+$0500 & $24.1$ & $\pm2.2$ & 5.0 & J1909$+$0007 & $-7.6$ & $\pm2.1$ & 1.4 & J1909$+$0912 & $119$ & $\pm31$ & 1.5 \\
J1910$+$0517 & $272$ & $\pm90$ & 1.2 & J1910$+$0728 & $-29.0$ & $\pm4.3$ & 1.9 & J1913$+$1011 & $-8.1$ & $\pm2.0$ & 0.91 \\
J1913$+$1145 & $-118$ & $\pm24$ & 1.3 & J1913$+$1400 & $-5.5$ & $\pm1.8$ & 0.68 & J1914$+$0219 & $10.4$ & $\pm3.4$ & 1.3 \\
J1915$+$0838 & $-126$ & $\pm32$ & 1.6 & J1916$+$0844 & $40.4$ & $\pm8.8$ & 1.0 & J1917$+$1353 & $12.46$ & $\pm0.86$ & 3.0 \\

\hline
\end{tabular}
\end{center}
\end{table*}

%% file: sigrm.tex
\begin{table*}
\caption{Pulsars for which the measured slope in RM is more than 3--$\sigma$ away from zero. Reduced $\chi^2$ of the linear fit is also given as an indication of the goodness of fit.}
\label{tab:sigrm}
\begin{center}
\begin{tabular}{crll|crll|crll}
\hline
\hline
PSR & \multicolumn{2}{c}{RM slope}  & $\chi^2$ & PSR & \multicolumn{2}{c}{RM slope} & $\chi^2$ & PSR & \multicolumn{2}{c}{RM slope}& $\chi^2$\\
 & \multicolumn{2}{c}{(rad\,m$^{-2}$yr$^{-1}$)} & & & \multicolumn{2}{c}{(rad\,m$^{-2}$yr$^{-1}$)}  & & & \multicolumn{2}{c}{(rad\,m$^{-2}$yr$^{-1}$)} & \\
\hline
J0536$-$7543 & $0.336$ & $\pm0.067$ & 2.8 & J0624$-$0424 & $-0.35$ & $\pm0.12$ & 4.0 & J0627$+$0649 & $-0.262$ & $\pm0.053$ & 3.2 \\
J0646$+$0905 & $-0.306$ & $\pm0.063$ & 2.2 & J0904$-$7459 & $0.435$ & $\pm0.099$ & 2.0 & J0909$-$7212 & $0.302$ & $\pm0.087$ & 2.8 \\
J0942$-$5657 & $0.216$ & $\pm0.048$ & 1.4 & J0943$+$1631 & $-0.76$ & $\pm0.24$ & 3.2 & J1001$-$5507 & $0.63$ & $\pm0.12$ & 11 \\
J1015$-$5719 & $-0.43$ & $\pm0.10$ & 2.5 & J1016$-$5857 & $1.20$ & $\pm0.20$ & 0.64 & J1047$-$6709 & $0.227$ & $\pm0.063$ & 2.2 \\
J1048$-$5832 & $0.923$ & $\pm0.049$ & 2.9 & J1056$-$6258 & $0.152$ & $\pm0.051$ & 3.1 & J1057$-$5226 & $0.126$ & $\pm0.027$ & 1.5 \\
J1114$-$6100 & $-2.18$ & $\pm0.10$ & 1.4 & J1123$-$6102 & $0.274$ & $\pm0.092$ & 0.63 & J1239$-$6832 & $0.46$ & $\pm0.12$ & 1.7 \\
J1326$-$6700 & $0.348$ & $\pm0.073$ & 4.4 & J1338$-$6204 & $0.247$ & $\pm0.082$ & 1.3 & J1452$-$6036 & $-1.39$ & $\pm0.40$ & 0.54 \\
J1507$-$4352 & $0.198$ & $\pm0.058$ & 2.6 & J1543$-$0620 & $0.170$ & $\pm0.050$ & 2.2 & J1602$-$5100 & $0.442$ & $\pm0.050$ & 3.8 \\
J1617$-$4216 & $0.48$ & $\pm0.14$ & 0.82 & J1637$-$4642 & $1.22$ & $\pm0.23$ & 0.66 & J1646$-$6831 & $0.303$ & $\pm0.099$ & 9.8 \\
J1648$-$6044 & $0.39$ & $\pm0.12$ & 0.90 & J1651$-$4246 & $0.292$ & $\pm0.041$ & 4.5 & J1703$-$3241 & $0.123$ & $\pm0.034$ & 3.2 \\
J1704$-$5236 & $1.18$ & $\pm0.28$ & 0.18 & J1717$-$5800 & $0.72$ & $\pm0.22$ & 0.52 & J1723$-$3659 & $-0.441$ & $\pm0.086$ & 0.64 \\
J1741$-$0840 & $0.202$ & $\pm0.064$ & 9.9 & J1743$-$3153 & $-1.53$ & $\pm0.42$ & 0.83 & J1759$-$2205 & $-0.562$ & $\pm0.066$ & 1.1 \\
J1803$-$2137 & $-0.456$ & $\pm0.089$ & 1.3 & J1808$-$0813 & $0.181$ & $\pm0.061$ & 0.52 & J1809$-$1917 & $-0.429$ & $\pm0.058$ & 0.73 \\
J1809$-$2109 & $1.21$ & $\pm0.40$ & 0.69 & J1825$-$1446 & $1.82$ & $\pm0.12$ & 1.2 & J1832$-$0827 & $-1.96$ & $\pm0.11$ & 0.22 \\
J1833$-$0338 & $0.310$ & $\pm0.069$ & 1.7 & J1833$-$0827 & $-13.11$ & $\pm0.88$ & 1.7 & J1834$-$0426 & $0.110$ & $\pm0.036$ & 2.1 \\
J1836$-$1008 & $-0.262$ & $\pm0.087$ & 1.9 & J1841$-$0345 & $-0.302$ & $\pm0.082$ & 0.82 & J1852$-$0118 & $1.74$ & $\pm0.27$ & 1.2 \\
J1852$-$0635 & $0.161$ & $\pm0.048$ & 4.6 & J1853$-$0004 & $1.54$ & $\pm0.31$ & 3.2 & J1856$-$0526 & $-1.01$ & $\pm0.31$ & 0.43 \\
J1906$+$0912 & $3.5$ & $\pm1.2$ & 0.51 & J1908$+$0909 & $1.30$ & $\pm0.43$ & 0.50 & J1909$+$0007 & $3.34$ & $\pm0.50$ & 11 \\
J1913$-$0440 & $0.149$ & $\pm0.041$ & 3.0 & J1915$+$1009 & $0.268$ & $\pm0.070$ & 5.4 & J2037$+$1942 & $-0.69$ & $\pm0.20$ & 3.4 \\

\hline
\end{tabular}
\end{center}
\end{table*}